\documentclass[
prd
,showpacs,amssymb,superscriptaddress,aps
]{revtex4}
\usepackage{graphicx}
\usepackage{color}
\input{epsf}

\usepackage{amsmath,amssymb}
\usepackage{bm}
\usepackage{times}
\usepackage{ulem}

\newcommand{\dalm}{\kern1pt\vbox{\hrule height 0.9pt\hbox{\vrule width 0.9pt
\hskip 2.5pt\vbox{\vskip 5.5pt}\hskip 3pt\vrule width 0.3pt}\hrule height 0.3pt}
\kern1pt}

\begin{document}



\title{Neutron star asteroseismology and nuclear saturation parameter}

\author{Hajime Sotani}
\email{sotani@yukawa.kyoto-u.ac.jp}
\affiliation{Astrophysical Big Bang Laboratory, RIKEN, Saitama 351-0198, Japan}
\affiliation{Interdisciplinary Theoretical \& Mathematical Science Program (iTHEMS), RIKEN, Saitama 351-0198, Japan}


\date{\today}

\begin{abstract}
Adopting various unified equations of state (EOSs), we examine the quasinormal modes of gravitational waves from cold neutron stars. We focus on the fundamental ($f$-), 1st pressure ($p_1$-), and 1st spacetime ($w_1$-) modes, and derive the empirical formulae for the frequencies and damping rate of those modes. With the resultant empirical formulae, we find that the value of $\eta$, which is a specific combination of the nuclear saturation parameters, can be estimated within $\sim 30 \%$ accuracy, if the $f$-mode frequency from the neutron star whose mass is known would be observed or if the $f$- and $p_1$-mode frequencies would be simultaneously observed, even though this estimation is applicable only for the low-mass neutron stars. Additionally, we find that the mass and radius of canonical neutron stars can be estimated within a few per cent accuracy via the simultaneous observations of the $f$- and $w_1$-mode frequencies. We also find that, if the $f$-, $p_1$-, and $w_1$-mode frequencies would be simultaneously observed, the mass of canonical neutron stars can be estimated within $2\%$ accuracy, while the radius can be estimated within $1\%$ for the neutron star with $M\ge 1.6M_\odot$ or within $0.6\%$ for the neutron star with $M\ge 1.4M_\odot$ constructed with the EOS constrained via the GW170817 event. Furthermore, we find the strong correlation between the maximum $f$-mode frequency and the neutron star radius with the maximum mass, between the minimum $w_1$-mode frequency and the maximum mass, and between the minimum damping rate of the $w_1$-mode and the stellar compactness for the neutron star with the maximum mass. With these correlation, one may constrain the upper limit of the neutron star radius with the maximum mass if a larger $f$-mode frequency would be observed, the lower limit of the maximum mass if a smaller $w_1$-mode frequency would be observed, or the lower limit of the stellar compactness for the neutron star with the maximum mass if a smaller value of the $w_1$-mode damping rate would be observed.
\end{abstract}

%
\maketitle


\section{Introduction}
\label{sec:I}

Neutron stars are born via the supernova explosions, which occur at the last moment of massive stars' life. Since the circumstance associated with the neutron stars are quite extreme, the neutron stars must be one of the most suitable objects for probing the physics under extreme states. In fact, the density inside the stars easily exceeds the nuclear saturation density, $\rho_0=2.68\times 10^{14}$ g/cm$^3$, and the gravitational and magnetic fields become much stronger than those observed in our Solar System \cite{ST83}. Thus, via the observations of neutron star itself and/or the phenomena associated with the neutron stars, one would extract the aspect of such extreme states, which enables us to understand the physics in neutron stars. The discovery of the $2M_\odot$ neutron stars \cite{D10,A13,C20} is a good example. Owing to the information about the existence of massive neutron stars, one could exclude some of soft equations of state (EOSs), with which the expected maximum mass is less than the observed mass. As another example, the light bending due to the relativistic effect is important, i.e., the light propagating through the strong gravitational field can bend. Because of this effect, the light curve from a rotating neutron star with hot spot(s) can be modified, which mainly depends on the stellar compactness defined as the ratio of the stellar mass to radius (e.g., Refs. \cite{PFC83,LL95,PG03,PO14,SM18,Sotani20a}). In practice, the properties of millisecond pulsar, PSR J0030+0451, have been already estimated via the x-ray observation with the Neutron star Interior Composition Explorer (NICER) mission \cite{Riley19,Miller19}. When one will get the further constraints on the mass and radius of the other neutron stars in a similar way, one can constrain the EOS for neutron star matter.

In addition, the oscillation frequency of neutron stars must be another important information. This is because, since the oscillation frequency strongly depends on the interior properties of neutron stars, one could inversely extract the interior information through the observation of the frequency. This technique is known as asteroseismology, which is similar to well-established seismology on Earth or helioseismology on Sun. As an example for the neutron star asteroseismology, by identifying the quasiperiodic oscillations observed in the afterglow following the giant flares from soft gamma-ray repeaters, SGR 1900+14 and 1806-20, with the neutron star crustal torsional oscillations, the crust properties in neutron stars could be constrained (e.g., Refs. \cite{GNHL2011,SNIO2012,SIO2016}). On the other hand, the gravitational waves are definitely suitable for adopting asteroseismology on neutron stars. In practice, it has been proposed that the neutron star mass ($M$), radius ($R$), and EOS would be constrained, once the gravitational waves coming from the neutron star oscillations will be observed (e.g., Refs. \cite{AK1996,AK1998,STM2001,SH2003,SYMT2011,PA2012,DGKK2013,Sotani20b,Sotani20c}). Moreover, in order to understand the physics behind the gravitational wave signals found in the numerical simulation for core-collapse supernovae, the gravitational wave asteroseismology on the protoneutron stars newly born just after the supernova explosion has been also studied (e.g., Refs. \cite{FMP2003,FKAO2015,ST2016,SKTK2017,MRBV2018,TCPOF19,SS2019,ST2020}).

As for the gravitational wave asteroseismology, the studies on the cold neutron stars have been extensively done. Since the neutron structures (and also the gravitational wave frequencies) strongly depend on the EOS for neutron star matter, which is still unfixed, the derivation of a kind of universal relation between the gravitational wave frequencies and some stellar properties, which is independent of the EOS, must be important for considering the asteroseismology. This is because one could know the stellar properties via such a universal relation independently of the EOS uncertainties, once the gravitational waves would be observed. Actually, it has been shown that the fundamental ($f$-) mode frequency, which is an acoustic oscillations, can be written as a function of the stellar average density, $M/R^3$, while the spacetime ($w$-) mode frequency can be characterized well as a function of the stellar compactness, $M/R$, almost independently of the neutron star EOS \cite{AK1996,AK1998}. On the other hand, in our previous studies \cite{Sotani20b,Sotani20c} it is found that the small EOS dependence in the relation between the $f$-mode frequency and the stellar average density comes from the EOS dependence of the stellar model and the corresponding $f$-mode frequency at the avoided crossing between the $f$- and the 1st pressure ($p_1$-) mode frequencies. Then, we proposed an alternative empirical formula for the $f$-mode frequency as a function of $M/R^3$ and nuclear saturation parameter, with which the stellar model and corresponding $f$-mode frequency at the avoided crossing can be characterized. However, these previous studies have been done with the relativistic Cowling approximation, i.e., the metric perturbations are neglected during the fluid oscillations. So, in this study, we will do a similar study without the Cowling approximation, especially focusing on the $f$-, $p_1$, and $w_1$-mode gravitational waves. Consequently, we can derive not only the empirical formula similar to that in the previous study, but also that as a function of the stellar properties.

This paper is organized as follows. In Sec. \ref{sec:EOS}, we mention the unified EOS adopted in this study. In Sec. \ref{sec:Oscillation}, we calculate the complex eigenfrequencies of gravitational waves from the neutron stars constructed with various EOSs. With the resultant eigenfrequencies, we will derive the empirical formulae for the oscillation frequencies and damping rate as a function of nuclear saturation parameter and/or stellar properties, such as the stellar average density and compactness. We will also show how well the empirical formulae work for estimating the stellar properties. Finally, we conclude this study in Sec. \ref{sec:Conclusion}. Unless otherwise mentioned, we adopt geometric units in the following, $c=G=1$, where $c$ denotes the speed of light, and the metric signature is $(-,+,+,+)$.

\section{Unified EOS adopted in this study}
\label{sec:EOS}

As in Refs. \cite{AK1996,AK1998,Sotani20b,Sotani20c}, in this study we simply consider the non-rotating cold neutron stars composed of the perfect fluid, where the stellar models are constructed by integrating the Tolman-Oppenheimer-Volkoff equations, assuming an appropriate EOS. A number of EOSs have been proposed up to now, which is based on different nuclear interaction, nuclear model, and composition, while in this study we particularly adopt the unified realistic EOSs as in Ref. \cite{Sotani20c}. Namely, we consider only the EOSs, which are consistently constructed for both of the core and crust region of neutron stars with the same framework. The EOSs adopted in this study are listed in Table \ref{tab:EOS}, where the EOS parameters for each EOS, such as the incompressibility, $K_0$; the slope parameter of the nuclear symmetry energy, $L$; and an auxiliary parameter, $\eta$, defined by $\eta \equiv (K_0L^2)^{1/3}$ \cite{SIOO14}, are shown together with the maximum mass of the neutron star constructed with each EOS. The last column, $x_{\rm AC}$, denotes the square root of the normalized stellar average density given by 
\begin{equation}
  x = \left(\frac{M}{1.4M_\odot}\right)^{1/2}\left(\frac{R}{10\ {\rm km}}\right)^{-3/2}
\end{equation}
for the stellar model at the avoided crossing between the $f$- and $p_1$-mode frequencies discussed in the next section. We remark that the values of $x_{\rm AC}$ derived in this study is different from those shown in Ref. \cite{Sotani20c}, which comes from dropping the Cowling approximation in this study. Considering the constraints on $K_0$ and $L$ obtained via the terrestrial experiments, i.e., $K_0=230\pm 40$ MeV \cite{KM13} and $L=58.9\pm16$ MeV \cite{Li19}, the fiducial value of $\eta$ becomes $70.5\lesssim \eta \lesssim 114.8$ MeV.

\begin{table}
\caption{EOS parameters adopted in this study, $K_0$, $L$, and $\eta$. In addition, the maximum mass, $M_{\rm max}/M_\odot$, for the neutron star constructed with each EOS, and the square root of the normalized stellar average density at the avoided crossing between the $f$- and $p_1$-modes, $x_{\rm AC}$, are also listed.} 
\label{tab:EOS}
\begin {center}
\begin{tabular}{ccccccc}
\hline\hline
EOS & $K_0$ (MeV) & $L$ (MeV) & $\eta$ (MeV) & $M_{\rm max}/M_\odot$ & $x_{\rm AC}$  \\
\hline
DD2
 & 243 & 55.0  & 90.2  & 2.41 & 0.1554 & \\ 
Miyatsu
 & 274 &  77.1 & 118  & 1.95 & 0.1519 & \\ 
Shen
 & 281 & 111  & 151  &  2.17  & 0.1325 & \\  
FPS
 & 261 & 34.9 & 68.2  & 1.80  & 0.1924 & \\ 
SKa
 & 263 & 74.6 & 114 & 2.22 & 0.1473 & \\ 
SLy4
 & 230 & 45.9 &  78.5 & 2.05 & 0.1587 & \\ 
SLy9
 & 230 & 54.9 &  88.4 & 2.16 & 0.1599 & \\  
Togashi
 & 245  & 38.7  & 71.6 & 2.21  & 0.1667 &  \\ 
\hline \hline
\end{tabular}
\end {center}
\end{table}

The advantage for adopting the unified EOSs is that the neutron star properties for low density region and the properties of low-mass neutron stars can be characterized well with $\eta$. The mass and gravitational redshift of the low-mass neutron stars are written as a function of $\eta$ and the ratio, $u_c$, of the central density, $\rho_c$, to the nuclear saturation density, $\rho_0$, i.e., $u_c\equiv \rho_c/\rho_0$, which leads to that the stellar radius is also written as a function of $\eta$ and $u_c$ \cite{SIOO14}. Additionally, the properties for slowly rotating low-mass neutron stars are associated with $\eta$ \cite{SSB16}, while  the possible maximum mass of neutron stars is also discussed as a function of $\eta$ \cite{Sotani17,SK17}. Moreover, we have shown that the value of $x$ for low-mass neutron stars and $x_{\rm AC}$ can be written as a function of $\eta$ \cite{Sotani20b,Sotani20c}.

\begin{figure}[tbp]
\begin{center}
\includegraphics[scale=0.5]{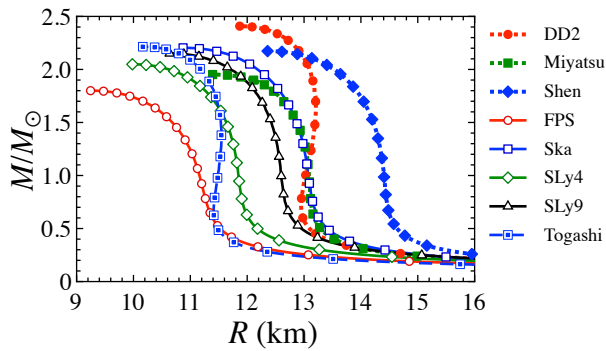}  
\end{center}
\caption{
Mass-radius relation for neutron star models constructed with various EOSs. In the figure, the dotted, solid, and dashed lines correspond to the EOSs based on the relativistic framework, the Skyrme-type effective interactions, and derived with variational method, respectively, while the marks on each line denote the background stellar models  for the linear analysis.
}
\label{fig:MR}
\end{figure}

In this study, as in Ref. \cite{Sotani20c} we adopt the EOSs based on the relativistic framework, i.e., DD2 \cite{DD2}, Miyatsu \cite{Miyatsu}, and Shen \cite{Shen}; the EOSs based on the Skyrme-type effective interaction, i.e., FPS \cite{FPS}, SKa \cite{SKa}, SLy4 \cite{SLy4}, and SLy9 \cite{SLy9}; and the EOS derived with a variational method, i.e., Togashi \cite{Togashi17}. In Fig. \ref{fig:MR}, the mass ($M$) and radius ($R$) relations for the neutron stars constructed with the EOSs adopted in this study are shown, where the marks on each line denote the neutron star models considered as a background model for the linear analysis. Although FPS and Shen EOSs may be excluded by the discovery of $2M_\odot$ neutron stars and the constraint from the gravitational wave event, GW 170817, i.e., the $1.4M_\odot$ neutron star radius should be smaller than 13.6 km \cite{Annala18}, in this study we adopt the both EOSs for examining the dependence on $\eta$ in wide range.

\section{Gravitational wave asteroseismology}
\label{sec:Oscillation}

On the neutron star models shown in the previous section, we consider to add linear perturbations. In the previous studies \cite{Sotani20b,Sotani20c} we simply adopted the relativistic Cowling approximation, where the metric perturbations are neglected during the fluid oscillations, while in this study we examine the linear analysis without the relativistic Cowling approximation, i.e., the metric perturbations are also taken into account. The perturbed variables and perturbation equations are concretely shown in Refs. \cite{STM2001,ST2020} together with the boundary conditions, where how to determine the frequency by solving the eigenvalue problem is also mentioned. As in the previous studies \cite{Sotani20b,Sotani20c}, we consider only the $\ell=2$ modes in this study, because it is considered that the $\ell=2$ modes are energetically more important in gravitational wave radiation.

By considering the metric perturbations, we have two advantages. First, since the gravitational waves would carry out the oscillation energy, the eigenfrequencies are complex, i.e., the so-called quasi-normal modes, where the real and imaginary parts correspond to the oscillation frequency and the damping rate, respectively. Thus, unlike the analysis with the Cowling approximation, now we can discuss the dependence of not only the oscillation frequencies but also the damping rate for each eigenmode. Second, one can observe not only the fluid oscillations, such as the $f$- and $p_i$-mode gravitational waves, but also the spacetime oscillations, such as the $w_i$-mode gravitational waves, which can be only discussed in the relativistic framework \cite{KS92}. It is known that the damping rate of the $w_i$-mode is generally very large, which is comparable to the oscillation frequency. In this study, we especially focus on the $f$-, $p_1$-, and $w_1$-modes, where the $p_1$- and $w_1$-modes are the eigenfrequencies with the lowest frequency in the $p_i$- and $w_i$-modes.

\begin{figure}[tbp]
\begin{center}
\includegraphics[scale=0.5]{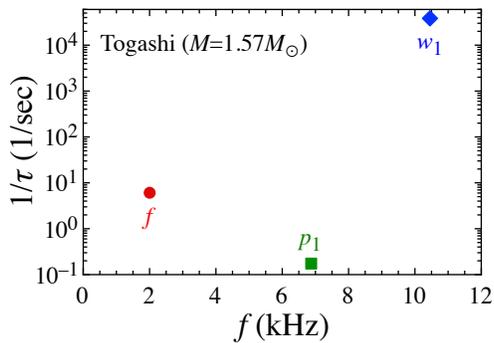}  
\end{center}
\caption{
Complex frequency (quasi-normal mode) for the neutron star model constructed with Togashi EOS, whose mass is $1.57M_\odot$. The real and imaginary parts of complex frequency correspond to the oscillation frequency, $f$, and damping rate, $1/\tau$, respectively. In particular, the $f$-, $p_1$-, and $w_1$-modes, which are discussed in this article, are shown here. 
}
\label{fig:Togashi13}
\end{figure}

\begin{figure}[tbp]
\begin{center}
\includegraphics[scale=0.5]{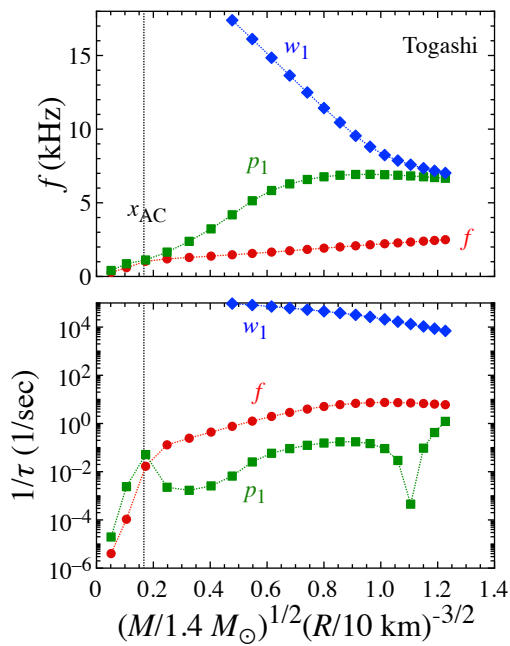}  
\end{center}
\caption{
The frequency, $f$, and damping rate, $1/\tau$, of the $f$-, $p_1$-, and $w_1$-modes for the neutron star models constructed with Togashi EOS are shown as a function of the square root of the normalized stellar average density, $x$, in the top and bottom panels, respectively. In the both panels, the value of $x_{\rm AC}$, where the avoided crossing between the $f$- and $p_1$-modes happens, is shown with the vertical line.
}
\label{fig:f-x-Togashi1}
\end{figure}

Once one selects the neutron star model, one can determine the complex value of the eigenfrequencies. For example, as shown in Fig. \ref{fig:Togashi13}, one can determine the $f$-, $p_1$-, and $w_1$-mode frequencies on the complex plane, where, with respect to the complex eigenvalue of $\omega$, we transform the values of Re$(\omega)$ and Im$(\omega)$ into Re$(\omega)=2\pi f$ and Im$(\omega)=1/\tau$ \cite{damping}. Then, as varying the stellar model with a specific EOS, the frequency and damping rate can be plotted as a function of $x$ in Fig. \ref{fig:f-x-Togashi1}. In the following, we will separately discuss the dependence of the frequency and damping rate.

\subsection{Oscillation frequency}
\label{sec:frequency}


Since the $f$-mode is a kind of the acoustic oscillations, it is known that the frequency can be expressed well as a function of $x$ \cite{AK1996,AK1998}. Even so, the small EOS dependence still remains (see the left panel of Fig. \ref{fig:ff-x1}), which mainly comes from the fact that the stellar model at the avoided crossing between the $f$- and $p_1$-mode frequency depends on EOS. In fact, by shifting $f_f$ with $f_{f,{\rm AC}}$, which is the $f$-mode frequency at the avoided crossing, and by shifting $x$ with $x_{\rm AC}$, we could drive the linear relation between $f_f-f_{f,{\rm AC}}$ and $x-x_{\rm AC}$ almost independently of the EOS \cite{Sotani20b,Sotani20c}. In order to derive the similar formula in this study, we start to determine the stellar model at the avoided crossing between the $f$- and $p_1$-mode frequencies.

\begin{figure*}[tbp]
\begin{center}
\includegraphics[scale=0.5]{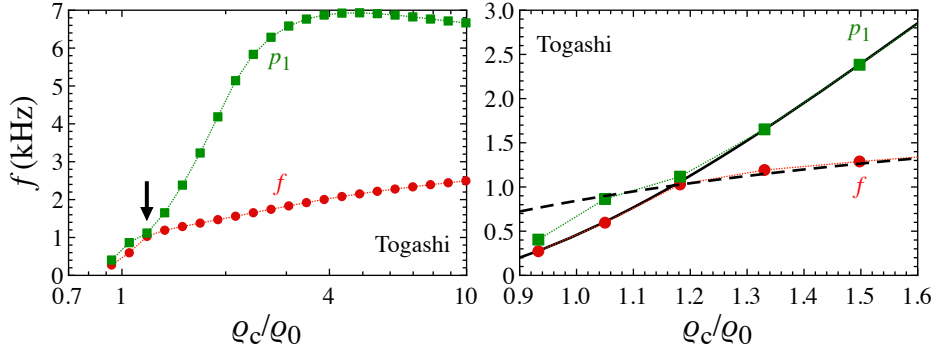}  
\end{center}
\caption{
In the left panel, as an example, the $f$- and $p_1$-mode frequencies for the neutron star model constructed with Togashi EOS are shown as a function of the central density, $\rho_c$, normalized by the saturation density, $\rho_0$. The arrow indicates the neutron star model at the avoided crossing between the $f$- and $p_1$-mode frequencies. In the right panel, the enlarged view is shown together with the fitting lines, where the solid and dashed lines correspond to the fitting given by Eqs. (\ref{eq:ff-uc1}) and (\ref{eq:ff-uc2}), respectively.
}
\label{fig:f-uc-Togashi}
\end{figure*}

In the previous studies with the Cowling approximation \cite{Sotani20b,Sotani20c}, in order to determine the central density, $\rho_{c, {\rm AC}}$, for the neutron star model at the avoided crossing between the $f$- and $p_1$-modes, the $f$-mode frequencies for the neutron star models with the central density, which is lower and higher than $\rho_{c, {\rm AC}}$, are respectively fitted by some function form with respect to the ratio, $u_c$, of the central density, $\rho_c$, to $\rho_0$. Then, $\rho_{c, {\rm AC}}$ is identified as the intersection of the fitting for the lower and higher density region. On the other hand, since the numerical cost for determining the eigenfrequencies is too much in this study, where the number of the stellar models is quite limited, it is difficult that one would determine the stellar model at the avoided crossing with the same way as in the previous studies. So, this time, the stellar model at the avoided crossing is determined as follows. That is, the $f$-mode frequencies for the stellar model with $\rho_c \le \rho_{c,{\rm AC}}$ and the $p_1$-mode frequencies for the stellar model with $\rho_c\ge \rho_{c,{\rm AC}}$ are fitted as a function of $u_c$ as Eq. (\ref{eq:ff-uc1}), while the $f$-mode frequencies for the stellar model with $\rho_c \ge \rho_{c,{\rm AC}}$ and the $p_1$-mode frequencies for the stellar model with $\rho_c\le \rho_{c,{\rm AC}}$ are fitted as a function of $u_c$ as Eq. (\ref{eq:ff-uc2}). The concrete fitting form is given by 
\begin{gather}
  f_f {\ \rm (kHz)} = a_1 + a_2 u_c + a_3 u_c^2 + a_4 u_c^3, \label{eq:ff-uc1} \\
  f_f {\ \rm (kHz)} = b_1 + b_2(\log u_c) + b_3(\log u_c)^2 + b_4(\log u_c)^3, \label{eq:ff-uc2}
\end{gather}
where $a_i$ and $b_i$ for $i=1-4$ are the adjusted coefficients depending on the EOS. Then, the intersection between Eqs. (\ref{eq:ff-uc1}) and (\ref{eq:ff-uc2}) is identified as the value of $\rho_{c,{\rm AC}}$ for the stellar model at the avoided crossing. As an example, in Fig. \ref{fig:f-uc-Togashi} we show the case with Togashi EOS. In the left panel, the $f$- and $p_1$-mode frequencies are shown as a function of $\rho_c/\rho_0$, where the stellar model at the avoided crossing is indicated with the arrow. The right panel is just an enlarged view of the left panel, where the fitting lines given by Eqs. (\ref{eq:ff-uc1}) and (\ref{eq:ff-uc2}) are also shown with the solid and dashed lines, respectively.

\begin{figure}[tbp]
\begin{center}
\includegraphics[scale=0.5]{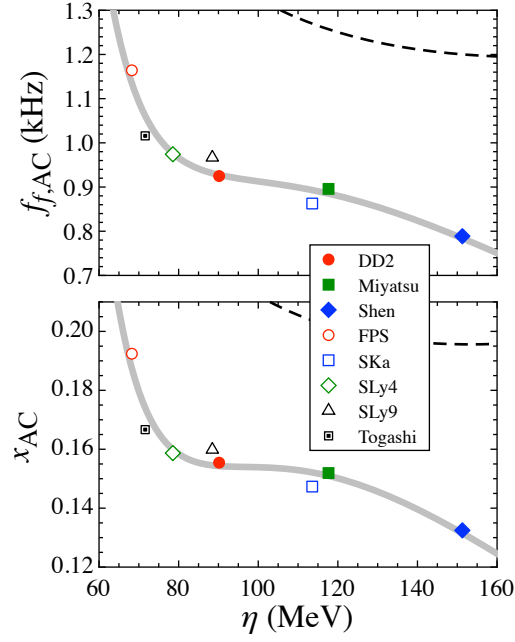}  
\end{center}
\caption{
The $f$-mode frequency, $f_{f,{\rm AC}}$, and the square root of the normalized stellar average density, $x_{\rm AC}$, for the neutron star model at the avoided crossing between the $f$- and $p_1$-mode frequencies are shown as a function of $\eta$ in the top and bottom panels, respectively. The thick-solid lines correspond to the fitting formulae given by Eqs. (\ref{eq:ff-ac}) and (\ref{eq:xac}). For reference, we also show the fitting lines with the Cowling approximation with the dashed lines, which correspond to Eqs. (A1) and (A2) in Ref. \cite{Sotani20c}.
}
\label{fig:ffac-xac}
\end{figure}

\begin{figure}[tbp]
\begin{center}
\includegraphics[scale=0.5]{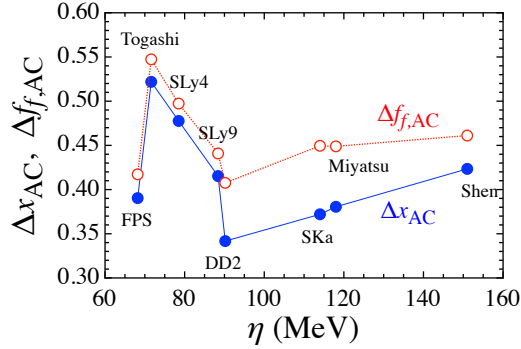}  
\end{center}
\caption{
Relative deviation of the value of $x_{\rm AC}$ and $f_{f,{\rm AC}}$ obtained in the previous study with the relativistic Cowling approximation \cite{Sotani20c} from those with full perturbation analysis in this study. The value of $\Delta x_{\rm AC}$ and $\Delta f_{f,{\rm AC}}$ are calculated with Eqs. (\ref{eq:DeltaX}) and (\ref{eq:Deltaff}), where the filled and open marks denote $\Delta x_{\rm AC}$ and $\Delta f_{f,{\rm AC}}$, respectively. 
}
\label{fig:Dxac}
\end{figure}

\begin{figure*}[tbp]
\begin{center}
\includegraphics[scale=0.45]{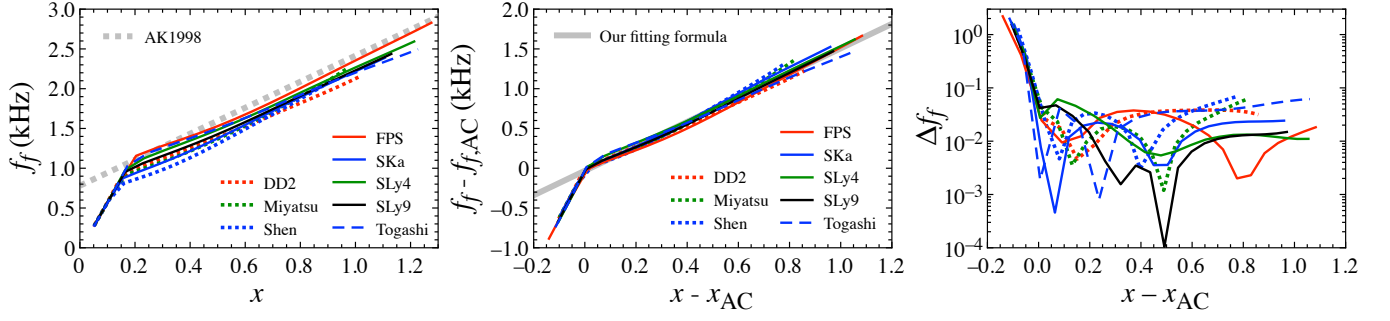}  
\end{center}
\caption{
In the left panel, the $f$-mode frequencies for various EOSs are shown as a function of the square root of the normalized average density. For reference, the empirical formula derived in Ref. \cite{AK1998} is shown with the thick-dotted line. In the middle panel, the shifted $f$-mode frequencies, $f_f-f_{f,{\rm AC}}$, are shown as a function of the shifted square root of the normalized average density, $x-x_{\rm AC}$, where the thick-solid line denotes the fitting formula given by Eq. (\ref{eq:ff-fit1}). In the right panel, the relative deviation, $\Delta f_f$, of the $f$-mode frequency estimated with Eqs. (\ref{eq:ff-fit2}) and (\ref{eq:ff-fit3}) from that calculated for each stellar model is shown as a function of $x-x_{\rm AC}$ for various EOSs, where $\Delta f_f$ is calculated with Eq. (\ref{eq:Delta_ff}).
}
\label{fig:ff-x1}
\end{figure*}

The value of $x_{\rm AC}$ is calculated for the stellar model with the central density of $\rho_{c,{\rm AC}}$ determined in such a way, while the value of $f_{f,{\rm AC}}$ is considered  in this study as the $f$-mode frequency, which is the intersection between Eqs. (\ref{eq:ff-uc1}) and (\ref{eq:ff-uc2}). In Fig. \ref{fig:ffac-xac}, for various EOSs, we show the values of $f_{f,{\rm AC}}$ in the top panel and $x_{\rm AC}$ in the bottom panel as a function of $\eta$. From this figure, one can see that the values of $f_{f,{\rm AC}}$ and $x_{\rm AC}$ are written as a function of $\eta$ as found in the previous studies \cite{Sotani20b,Sotani20c}, such as
\begin{gather}
  f_{f,{\rm AC}} {\ \rm (kHz)} = 1.8837\eta_{100}^{-3} -5.7685\eta_{100}^{-2} +  6.0096\eta_{100}^{-1} - 1.2120, \label{eq:ff-ac} \\
  x_{\rm AC} = 0.4126\eta_{100}^{-3} -1.2824\eta_{100}^{-2} +  1.3306\eta_{100}^{-1} - 0.3069, \label{eq:xac}
\end{gather}
where $\eta_{100}$ denotes the value of $\eta$ normalized by 100 MeV, i.e., $\eta_{100}\equiv \eta/ 100$ MeV. The expected values with these fitting formulae are also shown with the thick-solid lines in Fig. \ref{fig:ffac-xac}, where for reference the values with the fitting formulae obtained in the previous study with the Cowling approximation, i.e., Eqs. (A1) and (A2) in Ref. \cite{Sotani20c}, are also shown with dashed lines. Comparing the results with full perturbations to those with the Cowling approximation, we find that the avoided crossing with the Cowling approximation happens with larger value of $x$ and higher frequencies. In order to clarify this point, we calculate the relative deviation of the values with the Cowling approximation from those with full perturbations via 
\begin{gather}
  \Delta x_{\rm AC} = \frac{x_{\rm AC}^{\rm (Cow)} - x_{\rm AC}^{\rm (full)}}{x_{\rm AC}^{\rm (full)}}, \label{eq:DeltaX} \\
  \Delta f_{f,{\rm AC}} = \frac{f_{f,{\rm AC}}^{\rm (Cow)} - f_{f,{\rm AC}}^{\rm (full)}}{f_{f,{\rm AC}}^{\rm (full)}}, \label{eq:Deltaff}  
\end{gather}
where the properties calculated with the Cowling approximation and with full perturbations respectively denote with ``(Cow)" and ``(full)", and the resultant values are shown in Fig. \ref{fig:Dxac}. Namely, the values of $x_{\rm AC}$ and $f_{f,{\rm AC}}$ with the Cowling approximation are overestimated up to $\sim 50$ \%.

Now, as in the previous studies \cite{Sotani20b,Sotani20c}, we consider to derive the empirical formula for the $f$-mode frequency. In the left panel of Fig. \ref{fig:ff-x1}, we plot the $f$-mode frequencies for the neutron stars constructed with  various EOSs as a function of the square root of the normalized stellar average density, $x$, where the thick-dotted line denote the empirical formula derived in Ref. \cite{AK1998}. The bending point on each line at $x\simeq 0.1-0.2$ corresponds to the stellar model at the avoided crossing. It is obviously seen that the small EOS dependence for $x\ge x_{\rm AC}$ mainly  comes from the different bending point, depending on the EOS. In order to remove such EOS dependence, as shown in the middle panel of Fig. \ref{fig:ff-x1}, the $f$-mode frequencies shifted with $f_{f,{\rm AC}}$ is shown as a function of $x-x_{\rm AC}$. From this figure, one can observe that the value of $f_f-f_{f,{\rm AC}}$ for $x\ge x_{\rm AC}$ is well fitted as a linear function of $x-x_{\rm AC}$, such as
\begin{equation}
   f_f - f_{f,{\rm AC}}\ {\rm (kHz)} = 1.5390(x-x_{\rm AC}) - 0.03203. \label{eq:ff-fit1}
\end{equation}
The expected value with this fitting formula is also shown with the thick-solid line in the middle panel of Fig. \ref{fig:ff-x1}. Additionally, by substituting the fitting formulae of $f_{f,{\rm AC}}(\eta)$ and $x_{\rm AC}(\eta)$ given by Eqs. (\ref{eq:ff-ac}) and (\ref{eq:xac}) into Eq. (\ref{eq:ff-fit1}), one can obtain the empirical formula for the $f$-mode frequency as a function of $x$ and $\eta$ as
\begin{gather}
   f_f(x,\eta) \ {\rm (kHz)} = 1.5390x + f_0(\eta), \label{eq:ff-fit2} \\
   f_0(\eta)\ {\rm (kHz)} =  1.2487\eta_{100}^{-3} -3.7949\eta_{100}^{-2} +3.9618\eta_{100}^{-1} -0.7717.  \label{eq:ff-fit3}
\end{gather}
In order to assess the empirical formula, we calculate the relative deviation given by 
\begin{equation}
  \Delta f_f = \frac{|f_f - f_f(x,\eta)|}{f_f}, \label{eq:Delta_ff}
\end{equation}
where $f_f$ and $f_f(x,\eta)$ denote the $f$-mode frequency calculated as the eigenvalue problem for each stellar model and that estimated with the empirical formula, respectively, and the resultant value of $\Delta f_f$ is shown in the right panel of Fig. \ref{fig:ff-x1}. From this figure, we find that the empirical formula is valid even for the stellar model with the maximum mass within $\sim 10$ \% accuracy. For low-mass neutron stars, since $x$ can be expressed as a function of $\eta$ and $u_c$, i.e., $x=x(\eta,u_c)$, which is given by Eqs. (A8) - (A11) in Ref. \cite{Sotani20c}, one can rewrite the empirical formula for the $f$-mode frequency as a function of $u_c$ and $\eta$, i.e., $f_f=f_f(u_c,\eta)$.

\begin{figure*}[tbp]
\begin{center}
\includegraphics[scale=0.45]{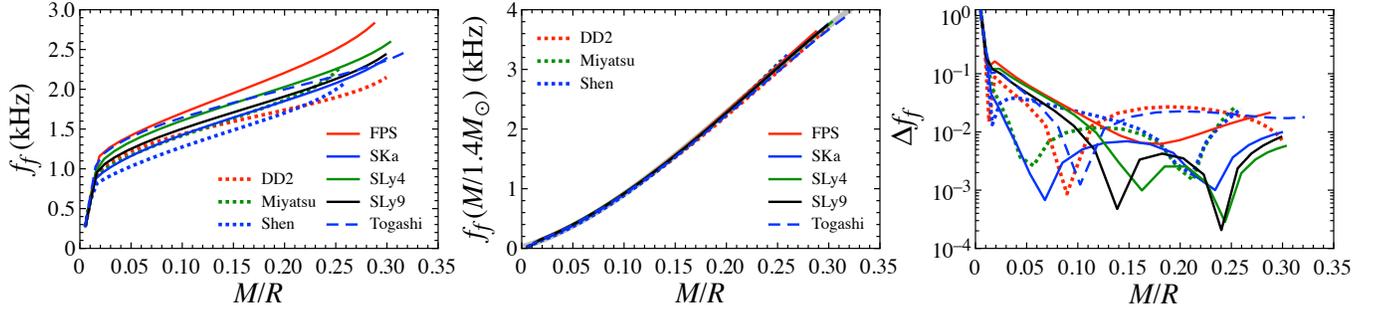}  
\end{center}
\caption{
In the left panel, the $f$-mode frequencies for various EOSs are shown as a function of $M/R$. In the middle panel, the $f$-mode frequencies multiplied by the normalized stellar mass, $f_f (M/1.4M_\odot)$, are shown as a function of $M/R$, where the thick-solid line denotes the fitting formula given by Eq. (\ref{eq:ffM-fit}). In the right panel, the relative deviation, $\Delta f_f$, of the $f$-mode frequency estimated with Eq. (\ref{eq:ffM-fit}) from that determined via the eigenvalue problem is shown as a function of $M/R$, where $\Delta f_f$ is calculated with Eq. (\ref{eq:Delta_ff}), using $f_f({\cal C},M)$ instead of $f_f(x,\eta)$.
}
\label{fig:ffM-x1}
\end{figure*}

In addition to the above empirical formula for the $f$-mode frequency, i.e., Eqs. (\ref{eq:ff-fit2}) and (\ref{eq:ff-fit3}), we find that the $f$-mode frequencies multiplied by the normalized stellar mass, $f_f(M/1.4M_\odot)$, are expressed as a function of the stellar compactness, ${\cal C}\equiv M/R$, independently of the EOS, as shown in Ref. \cite{TL05}. Actually, as shown in the left panel of Fig. \ref{fig:ffM-x1}, one can see the EOS dependence in the $f$-mode frequency as a function of $M/R$, while the EOS dependence in the value of $f_f(M/1.4M_\odot)$ as a function of $M/R$ almost disappears, as shown in the middle panel of Fig. \ref{fig:ffM-x1}. With using this behavior, we can derive the empirical formula for the $f$-mode frequency, such as
\begin{equation}
      f_{f}({\cal C},M)\ {\rm (kHz)} = \left[-66.3431 \left(\frac{M}{R}\right)^3 +43.9243 \left(\frac{M}{R}\right)^2 
        + 5.1709 \left(\frac{M}{R}\right) + 0.01847 \right]  \left(\frac{M}{1.4M_\odot}\right)^{-1}.   \label{eq:ffM-fit}
\end{equation}
In the right panel of Fig. \ref{fig:ffM-x1}, we show the relative deviation, $\Delta f_f$, of the $f$-mode frequency estimated with Eq. (\ref{eq:ffM-fit}) from that determined via the eigenvalue problem, which tells us that one can estimate the $f$-mode frequency from the canonical neutron stars within a few \% by using the empirical formula given by Eq. (\ref{eq:ffM-fit}). We remark that $M/R=0.172$ for the neutron star model with $M=1.4M_\odot$ and $R=12$ km.

\begin{figure*}[tbp]
\begin{center}
\includegraphics[scale=0.45]{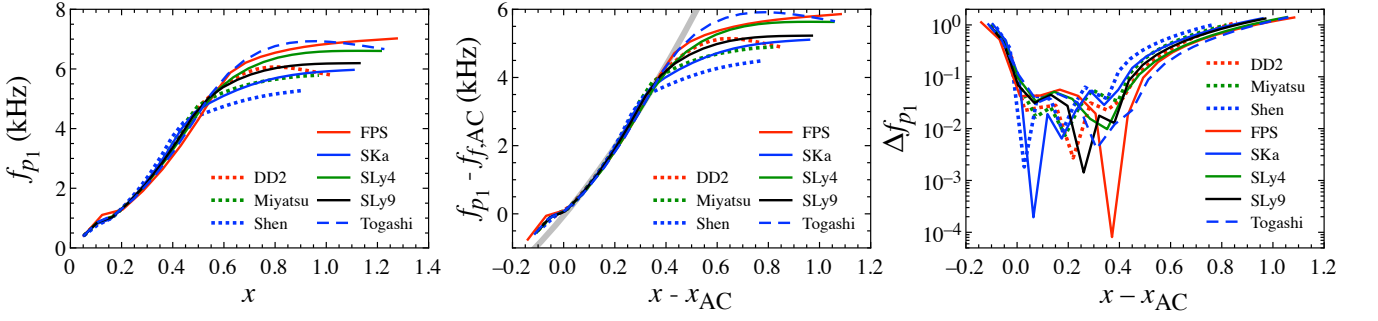}  
\end{center}
\caption{
In the left panel, the $p_1$-mode frequencies for various EOSs are shown as a function of $x$. In the middle panel, the shifted $p_1$-mode frequencies, $f_{p_1}-f_{f,{\rm AC}}$, are shown as a function of $x-x_{\rm AC}$, where the thick line denote the fitting formulae given by Eq. (\ref{eq:fp1-fit1}). In the right panel, the relative deviation of the $p_1$-mode frequencies for various EOSs are shown as a function of $x-x_{\rm AC}$, where $\Delta f_{p_1}$ is calculated with Eq. (\ref{eq:Delta_fp1a}).
}
\label{fig:fp1-x1}
\end{figure*}

Next, we consider the $p_1$-mode frequency. In the left panel of Fig. \ref{fig:fp1-x1}, we show the $p_1$-mode frequency as a functinon of $x$. Although it may be difficult to identify from this figure, three bending points exist on each line (c.f., Fig. 2 in Ref. \cite{Sotani20c}), i.e., the first is at $x\simeq 0.1$, the second is at $x\simeq 0.2$, and the third is at $x\simeq 0.5-0.7$, which respectively correspond to the stellar models at the avoided crossing between the $p_1$- and $p_2$-modes, between the $f$- and $p_1$-modes, and between the $p_1$- and $p_2$-modes again. Since the second avoided crossing in the $p_1$-mode frequency is strongly associated with the values of $f_{f,{\rm AC}}$ and $x_{\rm AC}$, we plot the $p_1$-mode frequency shifted with $f_{f,{\rm AC}}$ as a function of $x-x_{\rm AC}$ in the middle panel of Fig. \ref{fig:fp1-x1}. As discussed in Ref. \cite{Sotani20c}, one can see that the value of $f_{p_1} - f_{f,{\rm AC}}$ for the region between the second and third avoided crossing can be well fitted as a function of $x-x_{\rm AC}$ with
\begin{equation}
   f_{p_1} - f_{f,{\rm AC}}\ {\rm (kHz)} = 4.9701(x-x_{\rm AC})^2 +  9.0310(x-x_{\rm AC}) - 0.05404, \label{eq:fp1-fit1}
\end{equation}
with which the expected values are also shown in the middle panel of Fig. \ref{fig:fp1-x1} with the thick-solid line. By substituting the fitting formulae for $f_{f,{\rm AC}}(\eta)$ and $x_{\rm AC}(\eta)$ given by Eqs. (\ref{eq:ff-ac}) and (\ref{eq:xac}) into Eq. (\ref{eq:fp1-fit1}), the empirical formula for the $p_1$-mode frequency is expressed as 
\begin{equation}
   f_{p_1} \ {\rm (kHz)} = f_{p_1}(x,\eta), \label{eq:fp1-fit2}
\end{equation}
where we avoid to write down the full expression, because it would be too lengthy. In order to assess the empirical formula for the $p_1$-mode frequency, again we calculate the relative deviation given by
\begin{equation}
  \Delta f_{p_1} = \frac{|f_{p_1} - f_{p_1}(x,\eta)|}{f_{p_1}}, \label{eq:Delta_fp1a}
\end{equation}
where $f_{p_1}$ and $f_{p_1}(x,\eta)$ denote the $p_1$-mode frequency determined as the eigenvalue problem and that estimated with the empirical formula given by Eq. (\ref{eq:fp1-fit2}), respectively. The calculated value of $\Delta f_{p_1}$ is shown in the right panel of Fig. \ref{fig:fp1-x1}, which tells us that the value of the $p_1$-mode frequency for a neutron star model between the second and third avoided crossing in the $p_1$-mode frequency can be estimated with the empirical formula within $\sim 10$ \% accuracy. In a similar way to the $f$-mode frequency, using the relation of $x=x(\eta,u_c)$, one can rewrite the empirical formula for the $p_1$-mode frequency as a function of $u_c$ and $\eta$, i.e., $p_1=p_1(u_c,\eta)$, especially for low-mass neutron stars. We remark that the stellar model at the third avoided crossing in the $p_1$-mode frequency also depends on the adopted EOS (see the left panel of Fig. \ref{fig:fp1-x1}). So, if one could find a kind of a fitting formula expressing the stellar model at the third avoided crossing, such as $f_{f,{\rm AC}}(\eta)$ and $x_{\rm AC}(\eta)$ given by Eqs. (\ref{eq:ff-ac}) and (\ref{eq:xac}) for the avoided crossing between the $f$- and $p_1$-mode frequencies, one may derive a better empirical formula for the $p_1$-mode frequency even for the neutron star model above the third avoided crossing.

\begin{figure}[tbp]
\begin{center}
\includegraphics[scale=0.5]{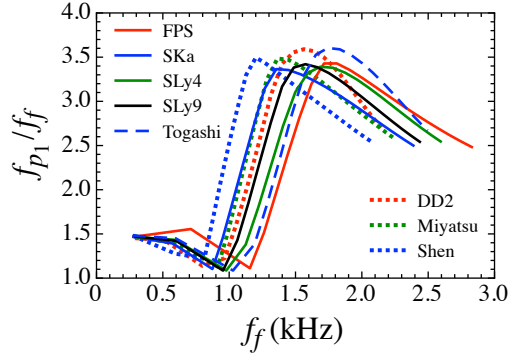}  
\end{center}
\caption{
Ratio of the $p_1$- to the $f$-mode frequency is shown as a function of the $f$-mode frequency for various EOSs. The minimum value of the ratio corresponds to the stellar model at the avoided crossing between the $f$- and $p_1$-mode frequencies, while the maximum value corresponds to that at the avoided crossing between the $p_1$- and $p_2$-mode frequencies. 
}
\label{fig:ff-fp1ff}
\end{figure}

With respect to the third avoided crossing in the $p_1$-mode frequency, we may say that the corresponding frequency, $f_{p_1,{\rm AC}3}$, depends on the EOS as shown in the left panel of Fig. \ref{fig:fp1-x1}, i.e., $f_{p_1,{\rm AC}3}\simeq 4-7$ kHz, while the ratio of the $p_1$- to the $f$-mode frequency, $f_{p_1}/f_f$, is almost independent of the EOS as shown in Fig. \ref{fig:ff-fp1ff}, i.e., $f_{p_1}/f_f\simeq 3.5$. In Fig. \ref{fig:ff-fp1ff}, the maximum value of $f_{p_1}/f_f$ correspond to the stellar model at the avoided crossing between the $p_1$- and $p_2$-mode frequencies, i.e., the third avoided crossing in the $p_1$-mode frequency, while the minimum value of $f_{p_1}/f_f$ is to that at the avoided crossing between the $f$- and $p_1$-mode frequencies, i.e., the second avoided crossing in the $p_1$-mode frequency, where by definition $f_{p_1}/f_f\simeq 1$.

\begin{figure*}[tbp]
\begin{center}
\includegraphics[scale=0.45]{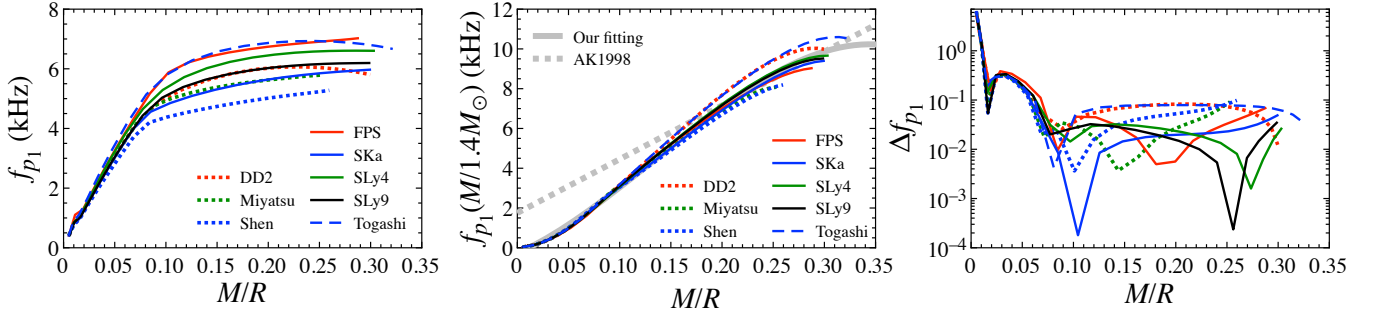}  
\end{center}
\caption{
In the left panel, the $p_1$-mode frequencies, $f_{p_1}$, are shown as a function of the stellar compactness, $M/R$, for various EOSs. In the middle panel, the $p_1$-mode frequencies multiplied by the normalized stellar mass, $f_{p_1}(M/1.4M_\odot)$, are shown as a function of $M/R$, where the thick-solid and thick-dotted lines denote the expectation with the fitting formula given by Eq. (\ref{eq:fp1-fit0}) and with the fitting derived in Ref. \cite{AK1998}, respectively. In the right panel, the relative deviation, $\Delta f_{p_1}$, of the $p_1$-mode frequency estimated with Eq. (\ref{eq:fp1-fit0}) from that calculated for each stellar model, where $\Delta f_{p_1}$ is calculated with Eq. (\ref{eq:Delta_fp1a}), using $f_{p_1}({\cal C},M)$ instead of $f_{p_1}(x,\eta)$.
}
\label{fig:fp1M}
\end{figure*}

On the other hand, we also derive the alternative empirical formula for the $p_1$-mode frequency. The $p_1$-mode frequency is shown as a function of $M/R$ in the left panel of Fig. \ref{fig:fp1M}, while the $p_1$-mode frequency multiplied by the normalized stellar mass, $f_{p_1}(M/1.4M_\odot)$, is shown in the middle panel. From this figure, one can obviously see the EOS dependence in $f_{p_1}$ as a function of $M/R$, while $f_{p_1}(M/1.4M_\odot)$ can be characterized well by $M/R$ with weak EOS dependence. This behavior is originally pointed out in Ref. \cite{AK1998}. In the middle panel of Fig. \ref{fig:fp1M}, we plot the estimation with the original fitting proposed in Ref. \cite{AK1998} with the thick-dotted line and that with the fitting obtained in this study given by 
\begin{equation}
   f_{p_1}({\cal C},M)\ {\rm (kHz)} = \left[-318.3935 \left(\frac{M}{R}\right)^3 +130.7236 \left(\frac{M}{R}\right)^2 
        + 23.2797 \left(\frac{M}{R}\right) - 0.2840 \right]  \left(\frac{M}{1.4M_\odot}\right)^{-1} \label{eq:fp1-fit0}
\end{equation}
with the thick-solid line. In order to assess this empirical formula for the $p_1$-mode frequency, we calculate the relative deviation with Eq. (\ref{eq:Delta_fp1a}), using $ f_{p_1}({\cal C},M)$ instead of $ f_{p_1}(x,\eta)$, and the resultant value of $\Delta f_{p_1}$ is shown in the right panel of Fig. \ref{fig:fp1M}. From this figure, we confirm that one can estimate the $p_1$-mode frequency from a canonical neutron star within $\sim 10$ \% accuracy, using the empirical formula given by Eq. (\ref{eq:fp1-fit0}).

\begin{figure*}[tbp]
\begin{center}
\includegraphics[scale=0.45]{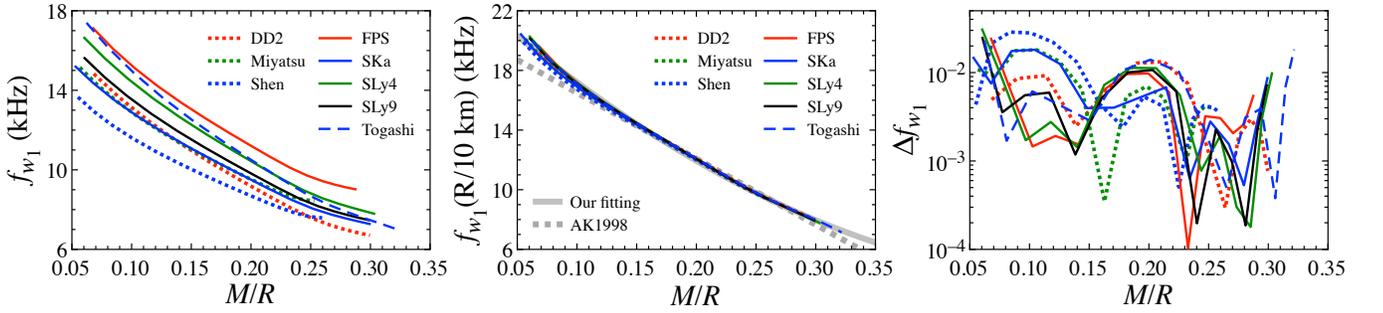}  
\end{center}
\caption{
In the left panel, the $w_1$-mode frequencies are shown as a function of the stellar compactness, $M/R$, for various EOSs. 
In the middle panel, the $w_1$-mode frequencies multiplied by the normalized stellar radius, $f_{w_1}(R/10\ {\rm km})$, are shown as a function of $M/R$, where the thick-solid and thick-dotted lines correspond to our fitting formula given by Eq. (\ref{eq:fw1-fit1}) and the fitting derived in Ref. \cite{AK1998}, respectively. In the right panel, the relative deviation, $\Delta f_{w_1}$, of the $w_1$-mode frequencies expected with Eq. (\ref{eq:fw1-fit1}) from that calculated for each stellar model is shown as a function of $M/R$, where $\Delta f_{w_1}$ is calculated with Eq. (\ref{eq:Delta_fw1}).
}
\label{fig:fw1R}
\end{figure*}

Finally, we consider the $w_1$-mode frequency. The fact that the $w_1$-mode frequency multiplied by the normalized stellar radius, $f_{w_1}(R/10\ {\rm km})$, can be expressed well as a function of $M/R$ has also been pointed out in Ref. \cite{AK1998}. In practice, the $w_1$-mode frequency is plotted as a function of $M/R$ in the left panel of Fig. \ref{fig:fw1R}, where one can still observe the EOS dependence, while $f_{w_1}(R/10\ {\rm km})$ can be characterized by $M/R$ almost independently of the EOS, as shown in the middle panel of Fig. \ref{fig:fw1R}. From this figure, we can derive the empirical formula for the $w_1$-mode frequency as
\begin{equation}
   f_{w_1}({\cal C},R) \ {\rm (kHz)} = \left[64.5819 \left(\frac{M}{R}\right)^2 - 72.0628 \left(\frac{M}{R}\right) + 23.7653\right]  
          \left(\frac{R}{\rm 10\ km}\right)^{-1}. \label{eq:fw1-fit1}
\end{equation}
In the middle panel of Fig. \ref{fig:fw1R}, the fitting formula given by Eq. (\ref{eq:fw1-fit1}) and that proposed in Ref. \cite{AK1998} are shown with the thick-solid and thick-dotted lines. Moreover, to see the accuracy of the empirical formula derived in this study, we check the relative deviation given by 
\begin{equation}
  \Delta f_{w_1} = \frac{|f_{w_1} - f_{w_1}({\cal C},R)|}{f_{w_1}}, \label{eq:Delta_fw1}
\end{equation}
where $f_{w_1}$ and $f_{w_1}({\cal C},R)$ denote the $w_1$-mode frequency determined by solving the eigenvalue problem and that estimated with Eq. (\ref{eq:fw1-fit1}), and show the value of $\Delta f_{w_1}$ in the right panel of Fig. \ref{fig:fw1R}. From this figure, we confirm that the $w_1$-mode frequency can be estimated with the empirical formula given by Eq. (\ref{eq:fw1-fit1}) within a few percent accuracy.

\begin{figure*}[tbp]
\begin{center}
\includegraphics[scale=0.5]{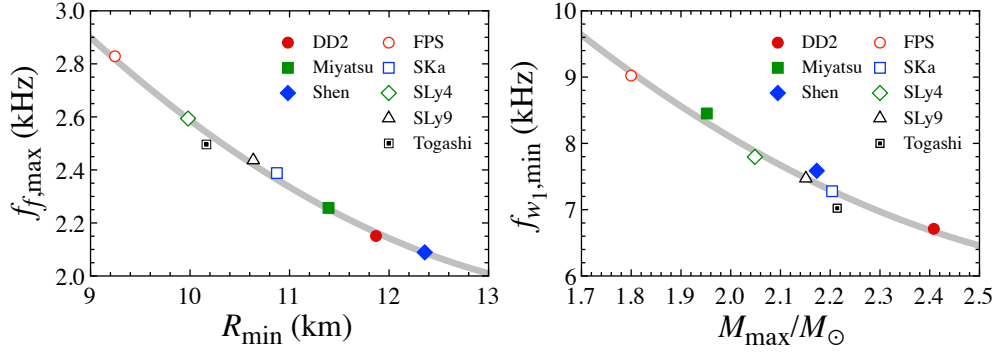}  
\end{center}
\caption{
In the left panel, the maximum frequency of the $f$-mode gravitational wave is shown as a function of the minimum radius of the neutron star constructed with each EOS. In the right panel, the minimum frequency of the $w_1$-mode gravitational wave is shown as a function of the maximum mass of the neutron star constructed with each EOS. The thick-solid lines in the left and right panels denote the fitting formulae given by Eqs. (\ref{eq:ffmax}) and (\ref{eq:fw1min}), respectively. 
}
\label{fig:ffmax}
\end{figure*}

Furthermore, in the previous study \cite{Sotani20c} we have shown that the maximum frequency of the $f$-mode gravitational wave is strongly correlated with the minimum radius, $R_{\rm min}$, of the neutron star, which corresponds to the neutron star with the maximum mass. With the $f$-mode frequency obtained in this study, we update such a correlation, i.e.,
\begin{equation}
  f_{f,{\rm max}}\ ({\rm kHz}) = 2.9799\left(\frac{R_{\rm min}}{10\ {\rm km}}\right)^2 
       - 8.7804 \left(\frac{R_{\rm min}}{10\ {\rm km}}\right) + 8.3878. \label{eq:ffmax}
\end{equation}
In the left panel of Fig. \ref{fig:ffmax}, the maximum $f$-mode frequency is shown as a function of $R_{\rm min}$ for various EOSs, where the fitting formula given by Eq. (\ref{eq:ffmax}) is also shown with the thick-solid line. In a similar way, we additionally find that the minimum $w_1$-mode frequency is strongly correlated with the maximum mass, $M_{\rm max}$, of the neutron stars, as shown in the right panel of Fig. \ref{fig:ffmax}, where the fitting formula is given by
\begin{equation}
  f_{w_1,{\rm min}}\ ({\rm kHz}) = 2.3459\left(\frac{M_{\rm max}}{M_\odot}\right)^2 
       - 13.8258 \left(\frac{M_{\rm max}}{M_\odot}\right) + 26.3605, \label{eq:fw1min}
\end{equation}
while we could not find any correlation between the maximum $p_1$-mode frequency and a specific neutron star property. Anyway, if one would observe a larger frequency of the $f$-mode gravitational wave, one could constrain the upper limit of $R_{\rm min}$ with Eq. (\ref{eq:ffmax}). If one would observe a smaller frequency of the $w_1$-mode gravitational wave, one could constrain the lower limit of $M_{\rm max}$ with Eq. (\ref{eq:fw1min}). Through the constraints on $R_{\rm min}$ and/or $M_{\rm max}$, one can narrow the EOS parameter space. That is, one can exclude some of the stiff EOSs with the maximum value of $R_{\rm min}$, while one can also exclude some of the soft EOSs with the minimum value of $M_{\rm max}$.

\subsection{Damping rate}
\label{sec:damping}

\begin{figure*}[tbp]
\begin{center}
\includegraphics[scale=0.45]{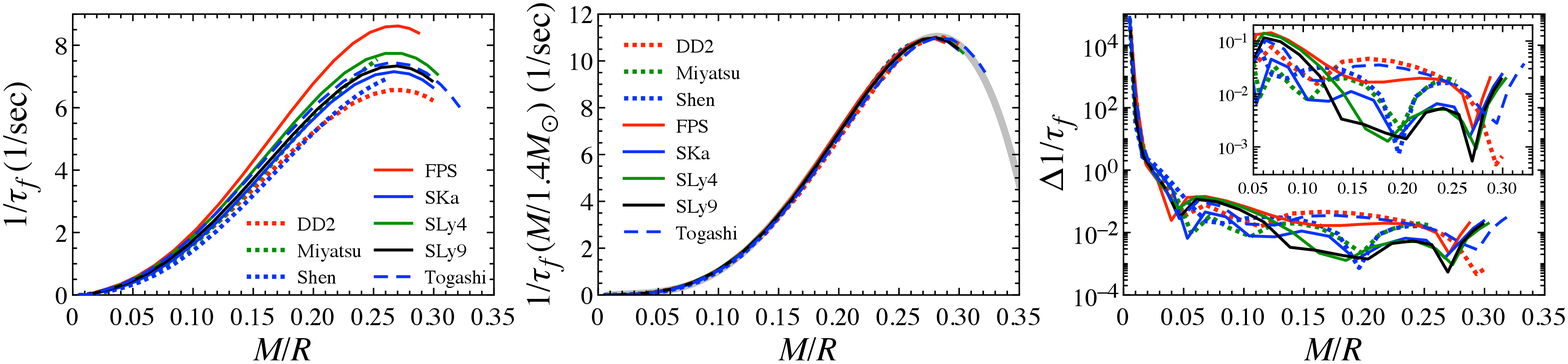}  
\end{center}
\caption{
In the left panel, the damping rate of the $f$-mode, $1/\tau_f$, is shown as a function of the stellar compactness, $M/R$, for various EOSs. In the middle panel, the damping rate of the $f$-mode multiplied by the normalized stellar mass, $1/\tau_f(M/1.4M_\odot)$, is shown as a function of $M/R$, where the thick-solid line corresponds to our fitting formula given by Eq. (\ref{eq:tauf-fit0}). In the right panel, the relative deviation, $\Delta 1/\tau_f$, of the damping rate of the $f$-mode expected with Eq. (\ref{eq:tauf-fit0}) from that calculated for each stellar model is shown as a function of $M/R$, where $\Delta 1/\tau_f$ is calculated with Eq. (\ref{eq:Delta_tauf}). The extended figure is also shown inside the right panel. 
}
\label{fig:tauf}
\end{figure*}

As mentioned before, owing to the analysis with metric perturbations, i.e., without the Cowling approximation, we can discuss not only the frequency but also the damping rate of the gravitational waves. The damping rate must be another information for extracting the neutron star properties, although the determination of the damping rate may be observationally more difficult than that of the frequency. First, the damping rate of the $f$-mode gravitational waves is shown as a function of $M/R$ in the left panel of Fig. \ref{fig:tauf}, where one can see a small EOS dependence. Meanwhile, as shown in the middle panel of Fig. \ref{fig:tauf}, we find that the damping rate of the $f$-mode multiplied by the normalized stellar mass, $1/\tau_f(M/1.4M_\odot)$, is expressed as a function of $M/R$ almost independently of the EOS, where the fitting formula is given by
\begin{equation}
   1/\tau_f({\cal C},M)\ {\rm (1/sec)} = \left[-7420.3659 \left(\frac{M}{R}\right)^4 +3290.3096 \left(\frac{M}{R}\right)^3 
        - 226.0861 \left(\frac{M}{R}\right)^2 + 7.8798 \left(\frac{M}{R}\right) - 0.06042 \right]  \left(\frac{M}{1.4M_\odot}\right)^{-1}. 
        \label{eq:tauf-fit0}
\end{equation}
In the right panel of Fig. \ref{fig:tauf}, in order to assess the empirical formula, the relative deviation, $\Delta 1/\tau_f$, between the damping rate, $1/\tau_f$, of the $f$-mode gravitational wave obtained by solving the eigenvalue problem and that estimated with the empirical formula given by Eq. (\ref{eq:tauf-fit0}), $1/\tau_f({\cal C},M)$, is shown as a function of $M/R$, where $\Delta 1/\tau_f$ is given by
\begin{equation}
  \Delta 1/\tau_f = \frac{|1/\tau_f - 1/\tau_f({\cal C},M)|}{1/\tau_f}. \label{eq:Delta_tauf}
\end{equation}
From this figure, we confirm that the damping rate of the $f$-mode gravitational waves from the canonical neutron stars can be estimated with the empirical formula given by Eq. (\ref{eq:tauf-fit0}) within $\sim 5$ \% accuracy.

\begin{figure}[tbp]
\begin{center}
\includegraphics[scale=0.5]{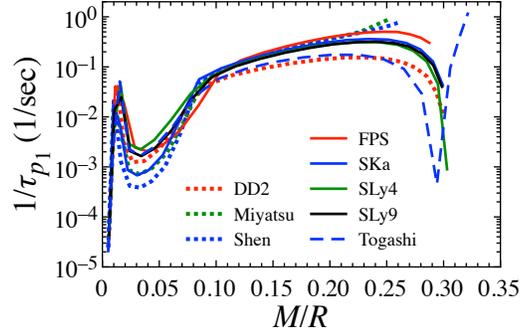}  
\end{center}
\caption{
The damping rate of the $p_1$-mode, $1/\tau_{p_1}$, is shown as function of the stellar compactness for various EOSs. 
}
\label{fig:taup1}
\end{figure}

Next, we consider the damping rate of the $p_1$-mode gravitational wave, but we can not get a kind of empirical formula. This may be because the $p_i$-mode has $i$ nodes in the eigenfunction, which leads to that the damping rate is quite sensitive to the stellar interior properties (or EOS). Moreover, as shown in Fig. \ref{fig:f-x-Togashi1}, the $p_1$-mode damping rate intersects with the $f$-mode damping rate, where the avoided crossing between the $f$- and $p_1$-mode frequency happens. With these reasons, the behavior of the $p_1$-mode damping rate seems to become quite complex, which strongly depend on the EOS, as shown in Fig. \ref{fig:taup1}.

\begin{figure*}[tbp]
\begin{center}
\includegraphics[scale=0.45]{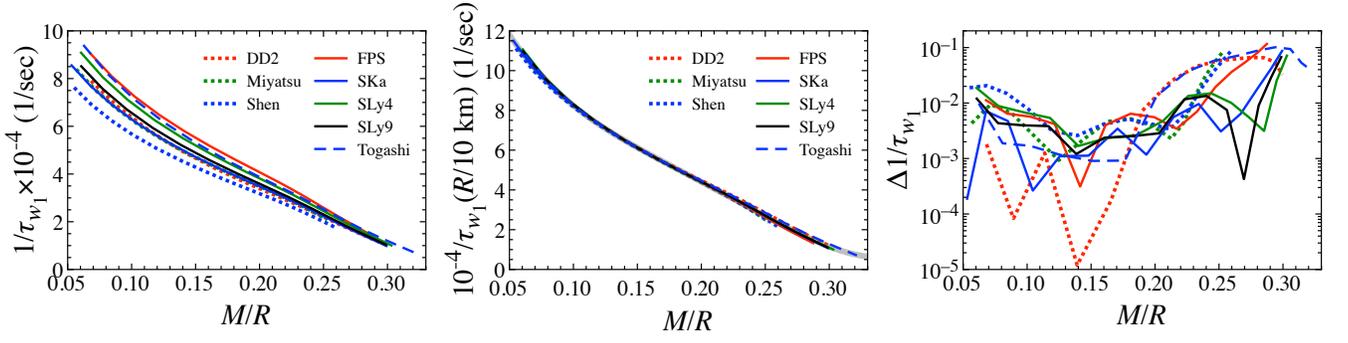}  
\end{center}
\caption{
In the left panel, the damping rate of the $w_1$-mode, $1/\tau_{w_1}$, is shown as a function of the stellar compactness, $M/R$, for various EOSs. In the middle panel, the damping rate of the $w_1$-mode multiplied by the normalized stellar radius, $1/\tau_{w_1}(R/10\ {\rm km})$, is shown as a function of $M/R$, where the thick-solid line corresponds to our fitting formula given by Eq. (\ref{eq:tauw1-fit}). In the right panel, the relative deviation, $\Delta 1/\tau_{w_1}$, of the damping rate of the $w_1$-mode expected with Eq. (\ref{eq:tauw1-fit}) from that calculated for each stellar model is shown as a function of $M/R$, where $\Delta 1/\tau_{w_1}$ is calculated with Eq. (\ref{eq:Delta_tauw1}). 
}
\label{fig:tauw1}
\end{figure*}

Finally, we consider the damping rate of the $w_1$-mode gravitational wave. In the left panel of Fig. \ref{fig:tauw1} we show the $w_1$-mode damping rate as a function of $M/R$, while in the middle panel of Fig. \ref{fig:tauw1} we show the $w_1$-mode damping rate multiplied by the normalized stellar radius, $f_{w_1}(R/10\ {\rm km})$. From this figure, one can obviously observe that $f_{w_1}(R/10\ {\rm km})$ can be expressed well as a function of $M/R$ independently of the EOS, which tells us the empirical formula for the $w_1$-mode damping rate, such as
\begin{equation}
   1/\tau_{w_1}({\cal C},R)\ {\rm (1/sec)} = \left[414.1431 \left(\frac{M}{R}\right)^4 - 344.8585 \left(\frac{M}{R}\right)^3 
       + 106.2514 \left(\frac{M}{R}\right)^2 - 17.7804 \left(\frac{M}{R}\right) + 1.8430 \right]  
       \left(\frac{R}{10\ {\rm km}}\right)^{-1} \times 10^5. 
        \label{eq:tauw1-fit}
\end{equation}
In the middle panel of Fig. \ref{fig:tauw1} we also show the values expected with the empirical formula with the thick-solid line. In the right panel of Fig. \ref{fig:tauw1}, we show the relative deviation, $\Delta 1/\tau_{w_1}$, calculated with
\begin{equation}
  \Delta 1/\tau_{w_1} = \frac{|1/\tau_{w_1} - 1/\tau_{w_1}({\cal C},R)|}{1/\tau_{w_1}}, \label{eq:Delta_tauw1}
\end{equation}
where $1/\tau_{w_1}$ and $1/\tau_{w_1}({\cal C},R)$ denote the $w_1$-mode damping rate determined via the eigenvalue problem and that estimated with the empirical formula, respectively. So, we confirm that the empirical formula for the $w_1$-mode damping rate can predict within $~10\%$ accuracy.

\begin{figure}[tbp]
\begin{center}
\includegraphics[scale=0.5]{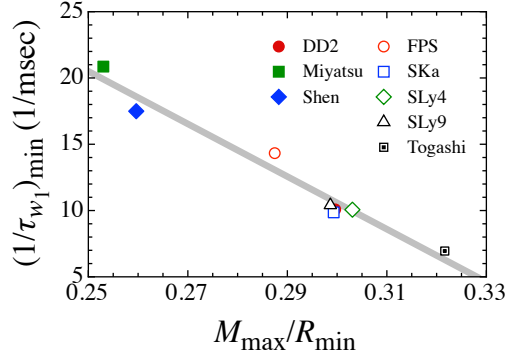}  
\end{center}
\caption{
The minimum damping rate of the $w_1$-mode gravitational wave is shown as a function of the stellar compactness for the neutron star with the maximum mass. The thick-solid line denotes the fitting formula given by Eq. (\ref{eq:tauwi_min}).
}
\label{fig:tauw1_min}
\end{figure}

Moreover, we find that the minimum value of the $w_1$-mode damping rate, which comes from the neutron star model with the maximum mass, is strongly associated with the compactness for the neutron star model with the maximum mass, $M_{\rm max}/R_{\rm min}$, as shown in Fig. \ref{fig:tauw1_min}, where the thick-solid line denotes the fitting formula given by 
\begin{equation}
  \left(1/\tau_{w_1}\right)_{\rm min}\ ({\rm 1/msec}) = -198.5094\left(\frac{M_{\rm max}}{R_{\rm min}}\right) + 70.1318. 
      \label{eq:tauwi_min}
\end{equation}
So, if one would observe a smaller value of $w_1$-mode damping rate, one could constrain the lower limit of $M_{\rm max}/R_{\rm min}$ via Eq. (\ref{eq:tauwi_min}).

At the end of this subsection, we mention about the possibility of damping by hydrodynamic effects. In this study we assume that the neutron star is composed of perfect fluid, but one may have to take into account the viscosity in reality. Nevertheless, since the damping time scale due to the viscosity would be in the order of years (e.g., Refs. \cite{CL87,Gusakov07}), the viscosity does not seriously affect at least the gravitational waves considered in this study. On the other hand, it may be crucial for considering the gravity ($g$-) mode oscillations, whose damping time can be comparable to the damping time scale due to the viscosity.

\subsection{Application}
\label{sec:application}

We collectively show the equation numbers corresponding to the  empirical formulae in Table \ref{tab:empi}, where we also list the mass formula, $M(u_c,\eta)$, for the low-mass neutron stars derived in Ref.  \cite{SIOO14}. Since it is considered that the frequency would be observationally determined easier than the damping rate, here we only focus on the empirical formulae for $f_f$, $f_{p_1}$, and $f_{w_1}$, and discuss how well those relations work for estimating the neutron star properties. In addition, since the stellar mass must be another observable, we also consider the case when the $f$-mode frequency would be observed from the neutron star, whose mass is known. In practice, in this section we discuss the possibility for determination of the value of $\eta$ with $M(u_c,\eta)$ and $f_f(u_c,\eta)$, or with $f_f(x,\eta)$ and $f_{p_1}(x,\eta)$. Then, we also discuss the possibility for determination of the stellar mass and radius with $f_f({\cal C},M)$ and $f_{w_1}({\cal C},R)$, with $f_{p_1}({\cal C},M)$ and $f_{w_1}({\cal C},R)$, or with $f_f({\cal C},M)$, $f_{p_1}({\cal C},M)$, and $f_{w_1}({\cal C},R)$.

\begin{table}
\caption{Correspondence of the equation numbers to the several empirical formulae. The equations derived in this study denote in boldface.} 
\label{tab:empi}
\begin {center}
\begin{tabular}{cc}
\hline\hline
observable & empirical formula    \\
\hline
$M(u_c,\eta)$     &  (2) in Ref. \cite{SIOO14}   \\
$f_f(x,\eta)$        & {\bf (\ref{eq:ff-fit2})}, {\bf (\ref{eq:ff-fit3})}   \\
$f_f(u_c,\eta)$    & {\bf (\ref{eq:ff-fit2})}, {\bf (\ref{eq:ff-fit3})} \& (A8) -- (A11) in Ref. \cite{Sotani20c} \\
$f_f({\cal C},M)$ & {\bf (\ref{eq:ffM-fit})}  \\
$f_{p_1}(x,\eta)$                & {\bf (\ref{eq:fp1-fit2})} or {\bf (\ref{eq:ff-ac})}, {\bf (\ref{eq:xac})}, {\bf (\ref{eq:fp1-fit1})}   \\
$f_{p_1}({\cal C},M)$         & {\bf (\ref{eq:fp1-fit0})}   \\
$f_{w_1}({\cal C},R)$         & {\bf (\ref{eq:fw1-fit1})}  \\
$1/\tau_f({\cal C},M)$         & {\bf (\ref{eq:tauf-fit0})}  \\
$1/\tau_{w_1}({\cal C},R)$ & {\bf (\ref{eq:tauw1-fit})}  \\
\hline \hline
\end{tabular}
\end {center}
\end{table}

\begin{figure}[tbp]
\begin{center}
\includegraphics[scale=0.5]{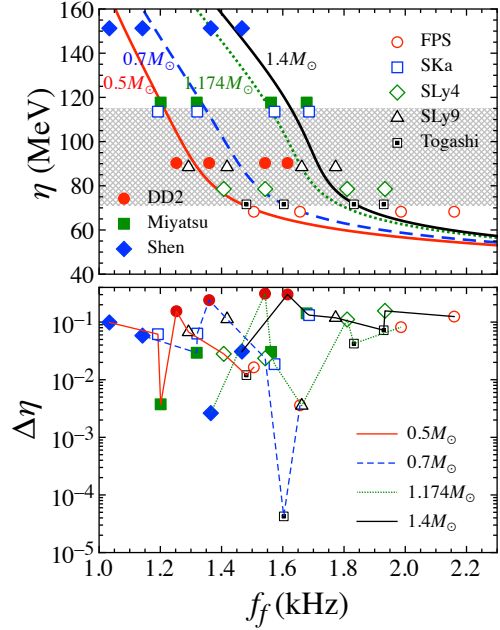}  
\end{center}
\caption{
In the top panel, with using the empirical formulae of $M(u_c,\eta)$ and $f_f(u_c,\eta)$, the expected value of $\eta$ is shown as a function of $f_f$ from the neutron star with $M/M_\odot=0.5$, 0.7, 1.174,  and 1.4 with various lines, where the lines from left to right correspond to the case with $M/M_\odot=0.5$, 0.7, 1.174,  and 1.4. The various marks denote the concrete numerical results obtained via the eigenvalue problem for various EOSs. For reference, the plausible value of $\eta$ constrained from the terrestrial experiments is shown by shaded region. In the bottom panel, the relative error, $\Delta\eta$, between the exact value of $\eta$ and the value of $\eta$ estimated with the empirical formulae is shown, when $f_f$ from the neutron star with the known mass would be observed. $\Delta\eta$ is calculated with Eq. (\ref{eq:deta}).
}
\label{fig:Deta-ffM}
\end{figure}

First, we consider the case that $f_f$ from the neutron star, whose mass is known, would be observed. Using the mass formula, $M(u_c,\eta)$, one would get the relation between $u_c$ and $\eta$, if the stellar mass is known. With this relation together with the formula of $f_f(u_c,\eta)$, one can estimate the value of $\eta$ as a function of $f_f$ for the given stellar mass. In the top panel of Fig. \ref{fig:Deta-ffM}, we show the value of $\eta$ expected from the empirical formulae for the neutron star model with $M/M_\odot=0.5$, 0.7, 1.174, and 1.4, where the exact value of $\eta$ for each EOS is also plotted with marks as a function of $f_f$ determined via the eigenvalue problem. Here, we remark that $1.174M_\odot$ is the smallest mass of neutron star observed up to now \cite{Martinez15}. To evaluate the accuracy of the estimation of $\eta$, we calculate the relative error given by
\begin{equation}
  \Delta\eta = \frac{|\eta - \eta_{\rm em}|}{\eta}, \label{eq:deta}
\end{equation}
where $\eta$ and $\eta_{\rm em}$ denote the exact value of $\eta$ and the value of $\eta$ estimated with the empirical formulae, and show the resultant values in the bottom panel of Fig. \ref{fig:Deta-ffM}. From this figure, we find that the value of $\eta$ can be estimated with the empirical formula within $\sim 30 \%$ accuracy, if the $f$-mode frequency of gravitational wave from the neutron star, whose mass is known and is less than $1.4M_\odot$, would be observed. This may be a surprising result, because the mass formula is originally derived only for the low-mass neutron stars, whose central density is less than twice the nuclear saturation density.

\begin{figure*}[tbp]
\begin{center}
\includegraphics[scale=0.5]{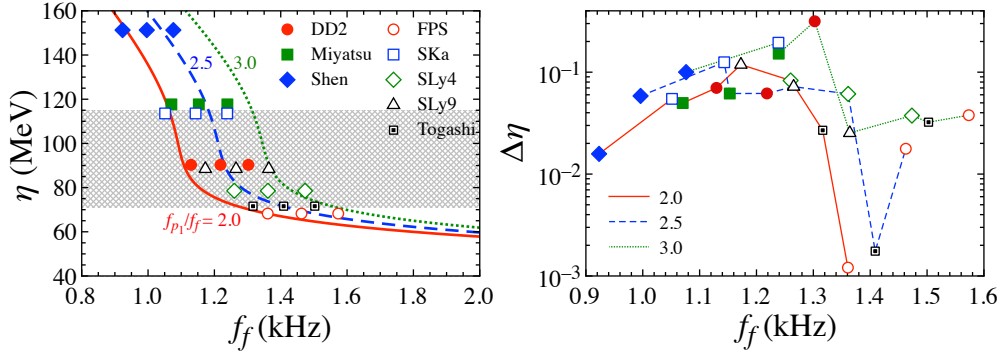}  
\end{center}
\caption{
In the left panel, with using the empirical formulae of $f_f(x,\eta)$ and $f_{p_1}(x,\eta)$, the estimated value of $\eta$ is shown as a function of $f_f$ for the case that the ratio of $f_{p_1}/f_f$ is constant, where the solid, dashed, and dotted lines denote the case with $f_{p_1}/f_f=2.0$, 2.5, and 3.0, respectively. The various marks denote the concrete value of $\eta$ with the numerical results determined via the eigenvalue problem. In the right panel, we show the relative error, $\Delta \eta$, is shown, where $\Delta\eta$ is calculated with Eq. (\ref{eq:deta}).
}
\label{fig:ff-eta-p1}
\end{figure*}

In a similar way, by using the empirical formulae of $f_f(x,\eta)$ and $f_{p_1}(x,\eta)$, we consider the possibility for estimating the value of $\eta$ via the simultaneous observation of $f_f$ and $f_{p_1}$. In the left panel of Fig. \ref{fig:ff-eta-p1}, we show the estimated value of $\eta$ as a function of $f_f$ for the case that the ratio of $f_{p_1}/f_f$ is constant, where the solid, dashed, and dotted lines dente the case with $f_{p_1}/f_f=2.0$, 2.5, and 3.0, respectively, while the exact value of $\eta$ for each EOS is also plotted with various marks, using the values of $f_f$ and $f_{p_1}$ determined via the eigenvalue problem. In the right panel of Fig. \ref{fig:ff-eta-p1}, we show the relative error calculated with Eq. (\ref{eq:deta}) for each EOS, which tells us that the value of $\eta$ can be estimated within $\sim 30\%$ accuracy with the empirical formula of $f_f(x,\eta)$ and $f_{p_1}(x,\eta)$. However, as shown in Fig. \ref{fig:fp1-x1}, since the empirical formula for the $p_1$-mode frequency, $f_{p_1}(x,\eta)$, can be adopted only for the low-mass neutron star, whose mass is less than that for the neutron star model at the avoided crossing between the $p_1$- and $p_2$-modes, the estimation discussed here may not so reasnable.

\begin{figure}[tbp]
\begin{center}
\includegraphics[scale=0.5]{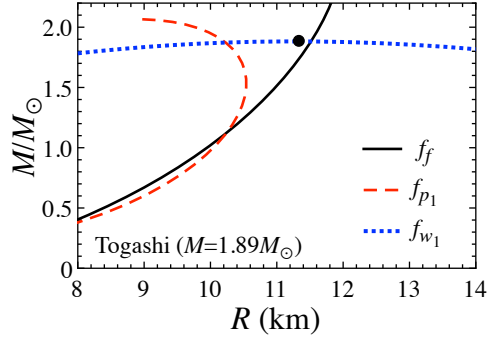}  
\end{center}
\caption{
The constraint on the relation between the mass and radius, given by $f_f({\cal C},M)={\rm const.}$ (solid line), $f_{p_1}({\cal C},M)={\rm const.}$ (dashed line), and $f_{w_1}({\cal C},R)={\rm const.}$ (dashed line) for a specific stellar model constructed with Togashi EOS with $1.89M_\odot$. The mark denotes the exact value of the corresponding mass and radius.
}
\label{fig:Togashi-M189}
\end{figure}

Next, we examine the possibility for estimation of the neutron star mass and radius via the observation of gravitational waves.  Using the empirical formulae of $f_f({\cal C},M)$, $f_{p_1}({\cal C},M)$, and $f_{w_1}({\cal C},R)$, one can get the relation between the mass and radius, if one would observe the $f$-, $p_1$-, or $w_1$-mode frequency of gravitational waves. As an example, in Fig. \ref{fig:Togashi-M189} we show the mass and radius relation obtained from $f_f({\cal C},M)$, $f_{p_1}({\cal C},M)$, and $f_{w_1}({\cal C},R)$, assuming that the $f$-, $p_1$-, and $w_1$-mode frequencies would be the specific values determined via the eigenvalue problem for the neutron star model with $M=1.89M_\odot$ constructed with Togashi EOS, where the exact value of mass and radius for the corresponding neutron star model is also plotted with the mark. So, for example, if one would simultaneously observe the $f$- and $w_1$-mode frequencies of gravitational wave, one can estimate the mass and radius of the source object as the intersection between the solid and dotted lines.

\begin{figure*}[tbp]
\begin{center}
\includegraphics[scale=0.45]{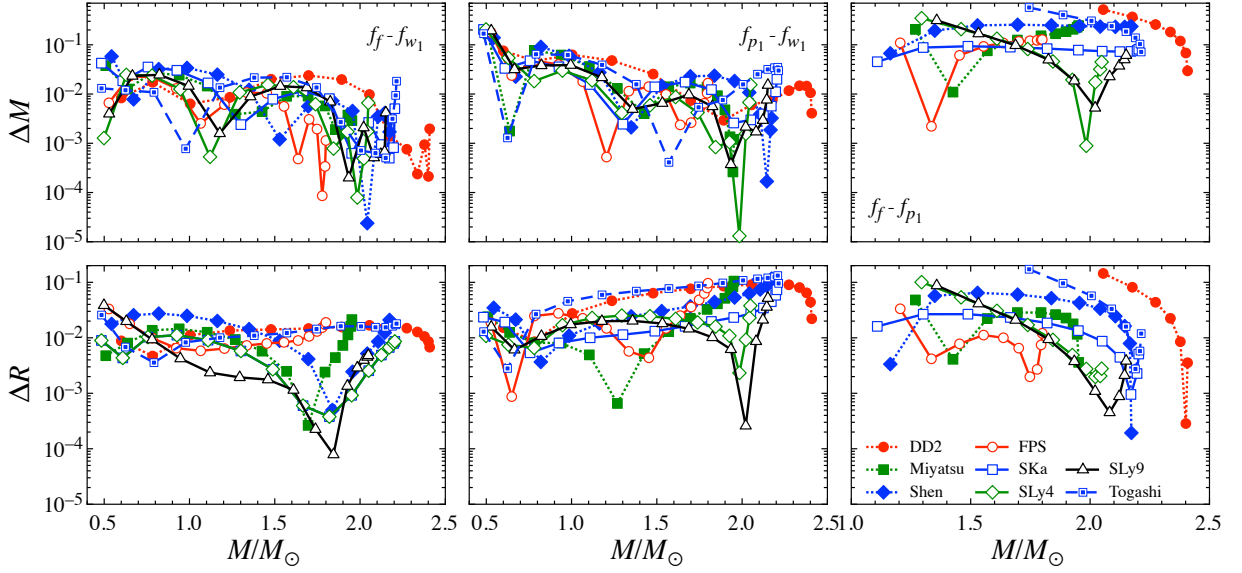}  
\end{center}
\caption{
With using the empirical formulae of $f_f({\cal C},M)$, $f_{p_1}({\cal C},M)$, and $f_{w_1}({\cal C},R)$, the relative error in mass, $\Delta M$, and in radius, $\Delta R$, between the exact value and the value expected with the empirical formulae is shown for various EOSs as a function of stellar mass, when $f_f$ and $f_{w_1}$ (left panel), $f_{p_1}$ and $f_{w_1}$ (middle panel), and $f_f$ and $f_{p_1}$ (right panel) would be simultaneously observed. $\Delta M$ and $\Delta R$ calculated with Eqs. (\ref{eq:dM}) and (\ref{eq:dR}), respectively.
}
\label{fig:DM-DR}
\end{figure*}

In order to assess the stellar mass and radius estimated in such a way with the empirical formula, we calculate the relative error given by 
\begin{gather}
  \Delta M = \frac{|M - M_{\rm em}|}{M}, \label{eq:dM} \\
 \Delta R = \frac{|R - R_{\rm em}|}{R}, \label{eq:dR} 
\end{gather}
where $M$ and $R$ denote the exact value of the mass and radius of the neutron star model for each EOS, while $M_{\rm em}$ and $R_{\rm em}$ denote the mass and radius estimated with the empirical formula when the gravitational wave frequency would be observed. In Fig. \ref{fig:DM-DR}, we show $\Delta M$ (top panels) and $\Delta R$ (bottom panels) as a function of the neutron star mass for various EOSs, where the left, middle, and right panels respectively correspond to the case when $f_f$ and $f_{w_1}$ would be simultaneously observed, when $f_{p_1}$ and $f_{w_1}$ would be simultaneously observed, and $f_f$ and $f_{p_1}$ would be simultaneously observed. From this figure, we find that one can estimate the mass and radius of canonical neutron stars within a few per cent via the observation of $f_f$ and $f_{w_1}$, while one can also estimate the mass within $4 \%$ and the radius within $15\%$ via the observation of $f_{p_1}$ and $f_{w_1}$. On the other hand, the estimation of neutron star mass and radius with $f_f$ and $f_{p_1}$ does not work well. In fact, one can estimate only for the massive neutron stars as shown in the right panel of Fig. \ref{fig:DM-DR}, where the leftmost stellar model on the each EOS shown in this panel denotes the stellar model with the smallest mass, which can be estimated with the empirical formula of $f_f$ and $f_{p_1}$, and there is no intersection between the mass and radius relation constrained via $f_f({\cal C},M)$ and $f_{p_1}({\cal C},M)$ for the stellar model with smaller mass. This is because both of the $f$- and $p_1$-modes are acoustic oscillations and the dependence of them on the stellar properties are very similar to each other.

\begin{figure}[tbp]
\begin{center}
\includegraphics[scale=0.5]{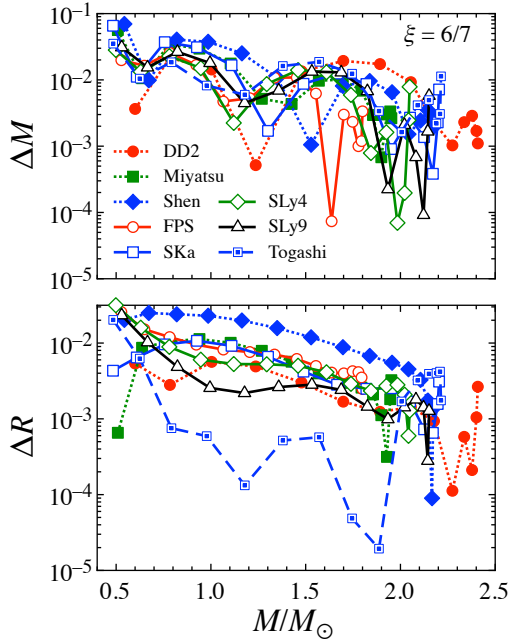}  
\end{center}
\caption{
The relative error in mass, $\Delta M$, and in radius, $\Delta R$, with using the estimations given by Eqs. (\ref{eq:dM}) and (\ref{eq:dR}), assuming that $\xi=6/7$. 
}
\label{fig:DM-DR-rate}
\end{figure}

Nevertheless, when one could observe the $p_1$- and $w_1$-mode frequencies, the $f$-mode frequency must be observed, because the $f$-mode frequency is located at the better frequency band in the sensitivity curve of the gravitational wave detectors rather than the $p_1$ and $w_1$-mode frequencies. In addition, as shown in Fig. \ref{fig:Togashi-M189}, the estimated radius with $f_f$ and $f_{w_1}$ (with $f_{p_1}$ and $f_{w_1}$) seems to be systematically larger (smaller) than the exact value of radius. So, here we propose the way to estimate the mass and radius of the neutron stars by using the mass and radius estimated via $f_f$ and $f_{w_1}$, which denote $M_{fw}$ or $R_{fw}$, and via $f_{p_1}$ and $f_{w_1}$, which denote $M_{pw}$ and $R_{pw}$, such as
\begin{gather}
  M_{\rm em} = \xi M_{fw} + (1-\xi) M_{pw}, \label{eq:Mem} \\
  R_{\rm em} = \xi R_{fw} + (1-\xi) R_{pw}. \label{eq:Rem}
\end{gather}
In practice, we find that the estimation of mass and radius becomes better especially for $\xi\simeq 0.83-0.88$. As an example, in Fig. \ref{fig:DM-DR-rate} we show the case with $\xi=6/7(=0.857)$. From this figure, we find that the mass of the canonical neutron stars are estimated within $2 \%$, while the radius of the neutron stars with $M\ge 1.6M_\odot$ is estimated within $1 \%$ through the simultaneous observations of the $f$-, $p_1$-, and $w_1$-mode frequencies. We also find that the radius of the neutron stars with $M\ge 1.4M_\odot$ can be estimated within $0.6\%$, if we neglect the results with Shen EOS, which has been already excluded via the gravitational wave observation from the binary neutron star merger event, GW170817 \cite{Annala18}.

\section{Conclusion}
\label{sec:Conclusion}

We examine the complex frequencies (quasi-normal modes) of gravitational waves, especially focusing on the $f$-, $p_1$-, and $w_1$-modes, from the neutrons stars constructed with various unified EOSs. First, as shown in the previous studies \cite{Sotani20b,Sotani20c}, we confirm that the $f$-mode frequency, $f_{f,{\rm AC}}$, and the square root of the normalized average density, $x_{\rm AC}$, of the neutron star at the avoided crossing between the $f$- and $p_1$-mode frequencies can be characterized well with a specific combination of the nuclear saturation parameters, $\eta$, where we also find that the avoided crossing appears with lower $f$-mode frequency and smaller stellar average density, compered to those with the Cowling approximation. In fact, the values of $f_{f,{\rm AC}}$ and $x_{\rm AC}$ are overestimated up to $\sim 50\%$ with the Cowling approximation. With using the relation of $f_{f,{\rm AC}}(\eta)$ and $x_{\rm AC}(\eta)$, we derive the empirical formula for the $f$- and $p_1$-mode frequencies as a function of $\eta$ and the square root of the normalized stellar average density, $x$, i.e., $f_f(x,\eta)$ and $f_{p_1}(x,\eta)$. These empirical formulae can be rewritten as a function of $u_c$ (the ratio of the stellar central density to the nuclear saturation density) and $\eta$, using the empirical formula of $x$ as a function of $\eta$ and $u_c$ derived in Ref. \cite{Sotani20c}. Then, we find that one can estimate the value of $\eta$ within $\sim 30 \%$ accuracy, if one would observe the $f$-mode frequency from the neutron star whose mass is known, adopting the mass formula derived in Ref. \cite{SIOO14}, i.e., $M(u_c,\eta)$, or if one would simultaneously observe the $f$- and $p_1$-mode frequencies. Even so, since the empirical formulae, $f_{p_1}(x,\eta)$ and $M(u_c,\eta)$, are applicable only for the low-mass neutron stars, the estimation of $\eta$ shown here may be possible in a limited occurrence.

On the other hand, we can derive alternative empirical formulae for the $f$-, $p_1$-, and $w_1$-mode frequencies and the damping rate of the $f$- and $w_1$-mode gravitational waves as a function of the stellar compactness, ${\cal C}\equiv M/R$,  and the mass (or radius) of the neutron stars independently of the EOS, i.e., $f_f({\cal C},M)$, $f_{p_1}({\cal C},M)$, $f_{w_1}({\cal C},R)$, $1/\tau_f({\cal C},M)$, and $1/\tau_{w_1}({\cal C},R)$. Owing to these empirical formulae, one might know the stellar mass and radius of the source object via the simultaneous observations of at least two of the five information in the gravitational waves, such as $f_f$, $f_{p_1}$, $f_{w_1}$, $1/\tau_f$, and $1/\tau_{w_1}$. As an example, we find that the mass and radius of canonical neutron stars can be estimated within a few per cent accuracy via the simultaneous observations of the $f$- and $w_1$-mode frequencies. We also find that, if the $f$-, $p_1$-, and $w_1$-mode frequencies would be simultaneously observed, the mass of canonical neutron stars can be estimated within $2\%$ accuracy, while the radius can be estimated within $1\%$ for the neutron stars with $M\ge 1.6M_\odot$ or within $0.6\%$ for the neutron stars with $M\ge 1.4M_\odot$ constructed with the EOS constrained via the GW170817 event.

Furthermore, we find that the maximum value of the ratio of the $p_1$- to the $f$-mode frequency is $\sim 3.5$ almost independently of the EOS. We also find the strong correlation between the maximum $f$-mode frequency and the neutron star radius with the maximum mass, between the minimum $w_1$-mode frequency and the maximum mass, and between the minimum damping rate of the $w_1$-mode and the stellar compactness for the neutron star with the maximum mass. With these correlation, one may constrain the upper limit of the neutron star radius with the maximum mass if a larger $f$-mode frequency would be observed, the lower limit of the maximum mass if a smaller $w_1$-mode frequency would be observed, or the lower limit of the stellar compactness for the neutron star with the maximum mass if a smaller value of the $w_1$-mode damping rate would be observed. These upper or lower limit on the neutron star properties must be crucial for constraining the EOS for dense matter.

\acknowledgments

This work is supported in part by Japan Society for the Promotion of Science (JSPS) KAKENHI Grant Numbers JP18H05236, JP19KK0354, JP20H04753, and 	JP21H01088  and by Pioneering Program of RIKEN for Evolution of Matter in the Universe (r-EMU).



\begin{thebibliography}{999}
\bibitem{damping}
   In Ref. \cite{ST2020}, the damping rate, $1/\tau$, has been calculated with $1/\tau={\rm Im}(\omega)/2\pi$, which is just mishandling in the calculation. In general, $1/\tau$ should be $1/\tau={\rm Im}(\omega)$ without $2\pi$ as in text.
\end{thebibliography}

\end{document}